\documentclass[twocolumn,showpacs,amsmath,amssymb,prb]{revtex4}
\usepackage{graphicx}
\usepackage{dcolumn}
\usepackage{bm}
\usepackage{amsmath}
\usepackage{amssymb}
\usepackage[latin1]{inputenc}
\usepackage{changebar}

\usepackage{mathptmx}

\newcounter{saveeqn}

\begin{document}

\title{Surface influence on stability and structure of III-V nanorods: \emph{First-principles} studies}

\author{R. Leitsmann and F. Bechstedt}
\affiliation{Institut f\"ur Festk\"orpertheorie und -optik,
Friedrich-Schiller-Universit\"at, Max-Wien-Platz 1, 07743 Jena,
 Germany }
\date{\today}
\begin{abstract}
We report \emph{ab initio} investigations of hexagon-shaped,
[111]/[0001] oriented III-V semiconductor nanowires with varying
crystal structure, surface passivation, surface orientation, and diameter. Their stability is dominated by the free surface
energies of the corresponding facets which differ only weakly from
those of free surfaces. We observe a phase transition between local zinc-blende
and wurtzite geometry versus preparation
conditions of the surfaces, which is accompanied by a change in the facet orientation. The influence of the actual III-V compound remains small. The atomic relaxation of nanowires gives rise to smaller bond lengths in comparison to the bulk znic-blende structures resulting
also in somewhat reduced bilayer thicknesses parallel to the growth
direction.
\end{abstract}
 \maketitle

\section{Introduction}

Highly anisotropic needlelike crystals (whiskers) have long been
subject of physics and materials science. This interest, especially
in the ultimately thin varieties, has been recently stimulated by the
potential need as building blocks for nanoscale electronic and
photonic devices \cite{1,2}. The synthesis of nanowires based on a
wide range of material systems \cite{3,4,5,6,7,8,9,10} have been
reported.
Even heterostructures have been prepared as vertically segmented wires along the growth axis or as core-shell structures perpendicular to the wire orientation.~\cite{A,B} 
Diameters varying from about one to hundred nanometers
and lengths extending to several micrometers are observed. Because
of their considerable technological potential for optoelectronics
or high-speed electronics, nanowires or nanorods consisting of III-V
semiconductors, AlAs, GaAs, InAs, GaP, and InP,
are of special interest. In the most cases the
growth direction of III-V semiconductor nanorods is parallel to
a [111] axis of the bulk zinc-blende ($zb$) substrate, which is
the ground-state structure of the common III-V compounds.

However,
in contrast to bulk crystals the crystal structure of the
nanowires may differ noticable, depending on growth conditions and
growth method. In particular, changes of the crystal symmetry from
cubic to hexagonal stacking of the cation-anion bilayers
have been observed in many cases \cite{B,3,4,7,8,9,10}. 
There are
reports of a rotational twin structure around the [111] growth
axis or even of the wurtzite ($w$) structure with [0001] stacking
direction~\cite{B,3,4} (schematically indicated in Fig.~1). 
This has been discussed in detail for GaAs and the unique vapor-liquid-solid growth mechanism.\footnote{Usually with a liquid metal seed particle and based on solid phase diffusion.}
The growth temperature has an strong influence in the case of the metal-organic vapor-phase epitaxy (MOVPE). Free-standing GaAs nanorods possess \emph{zb} structure for the temperature range 460-500$^{\text{o}}$C but changes to the \emph{w} polytype at 420$^{\text{o}}$C or above 500$^{\text{o}}$C.~\cite{2} In the case of the chemical beam epitaxy (CBE) the \emph{w} structure has been observed at 540$^{\text{o}}$C or below 400$^{\text{o}}$C.~\cite{C} Laser-assisted catalytic growth at higher growth temperatures usually results in the \emph{zb} phase.~\cite{6}

The consequences of the stacking variation for the geometry parameters such as bond lengths are not clarified. No significant changes have been reported for the bilayer thickness in [111] growth direction.~\cite{5,6,10}
Only a very small reduction of the distance between (111)
lattice planes compared to bulk values was observed in transmission electron
microscopy (TEM) studies for GaAs~\cite{5,6} and InP.~\cite{10} Other studies report a slight increase of the $c/a$ ratio discussing InAs in terms of a hexagonal geometry.~\cite{Mandl,Mandl19}
\begin{figure}
\includegraphics[height=5cm]{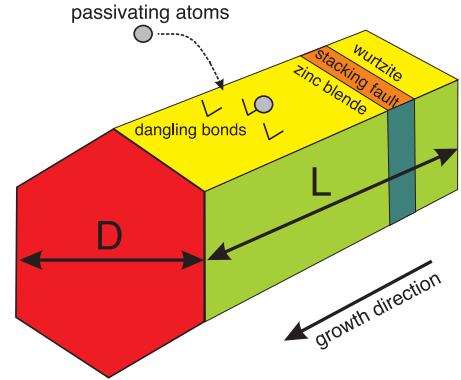}
\caption{(Color online) Schematic picture of a hexagon-shaped nanorod with diameter $D$ and a certain length $L$, which is a multiple of irreducible slabs.}
\end{figure}

Another not well clarified point concerns the favorable orientation of the surface facets. There seems to be agreement about the hexagon shape of the rod cross-sections with equivalent facet orientation.~\cite{3,7,8,10,15b,Mandl}
For GaAs rods few papers report the orientation of the side facets, either \{$11\bar 2$\} (\emph{zb}) / \{$1\bar 100$\} (\emph{w}) or \{$1\bar 10$\} (\emph{zb}) / \{$11\bar 20$\} (\emph{w}).~\cite{3,7} Scanning electron microscopy investigations found hexagonal cross-sections with \{$112$\} side facets for GaP nanorods with \emph{zb} stacking,~\cite{D} while \{$1\bar 10$\} side facets have been reported for InAs and InP nanorods.~\cite{Mandl,15b,10} However, two-dimensional defects such as twin segments or altering orientation may lead to wires that are microfaceted.~\cite{E} Growth of wires with triangular cross-sections may also occur as found for GaAs rods with \{$112$\} side facets grown in [111]A direction.~\cite{D}
Relatively little is known about the surface bonding and the contamination of the side facets. Because surface oxides such as In$_2$O$_3$ on InAs desorb above 500$^{\text{o}}$C, one thinks that annealing at lower temperatures results in incomplete desorption of the surface contamination.~\cite{4} Of course, growth in an ultrahigh vacuum system should result in almost contamination-free surfaces. However, also amorphous overlayers have been observed and attributed to oxide layers that form after GaAs is exposed to air.~\cite{5}  

The growth of quasi-one-dimensional III-V semiconductor rods
with varying crystal structure and bilayer distances raises the
question of the limit of a bulklike description of bonding in
these nanowires. Especially for small diameters the predominance
of surface atoms over inner, bulklike atoms will eventually lead
to bonds and, hence, geometries different from those of the bulk
systems. Thereby, the details of such changes may be influenced by the reconstruction/relaxation of the clean surfaces or the stoichiometry/thickness of the surface contamination. Despite varying numbers of dangling bonds and a possible
passivation of these dangling bonds, trials to understand the
surface influence are almost missing. The interplay of bonding
arrangement and stoichiometry of rod surfaces and $zb-w$ phase
transition has not been studied so far.  

On reason for this lack is the limitation of \emph{first principles} calculations to rather small (thin) systems. This hampers the comparability with experimental results. Parameter-free calculations were performed mainly on carbon or silicon nanowires/tubes. The investigated structures had diameters up to about one nanometer~\cite{6a,6b}, which is close to the thinest experimentally observed structures. However, to make reliable predictions for the $zb-w$ phase transition in III-V nanowires it is necessary to deal with larger systems.

In the present article, we try to overcome the limitations of \emph{ab initio} calculations by combining two approaches, microscopic total-energy studies of ultrathin nanorods with varying surface facet orientations, bonding and overlayers with macroscopic (thermodynamic) studies based on surface energies which are computed for individual surfaces. We demonstrate 
rapid convergence of the rod formation energy per surface area of III-V semiconductor nanowires with rising diameter. This fact enables us to extrapolate our results to diameters observed experimentally. In particular, we examine
free-standing, hexagon-shaped GaAs, InP, and InAs nanowires with
varying surface facet orientation and passivation, several
diameters, and two crystal structures close to the thermal
equilibrium. The results are extrapolated to the case of extremely
thick rods with almost bulklike surfaces. The favored geometries
are studied versus the preparation conditions of the wires.

\section{Modelling of Nanorods}
\subsection{Total energy calculation}

The total-energy (TE) calculations are performed in the framework
of the density functional theory (DFT) and the local density
approximation (LDA) as implemented in the Vienna \emph{ab initio}
simulation package (VASP) \cite{12a,12b}. The outermost $s$-, $p$-, and
(in the case of In) $d$-electrons are treated as valence
electrons whose interaction with the remaining ions is modeled by
pseudopotentials generated within the projector-augmented wave
(PAW) method \cite{13}. The energy cutoff of the plane-wave basis
is tested to be sufficient with 15 Ry. To model nanowires we use the common supercell approach~\cite{13b}. Neighboring rods are separated by a vacuum
region of about 2~nm to minimize artificial interactions across the periodic boundaries. The Brillouin-zone
summations are restricted to M$\times$N$\times$2 meshes of special
points according to Monkhorst and Pack~\cite{14}, with M$=$N$=$5
for diameters D (as defined in Fig.~1 as the vertical distance of two parallel surface facets) smaller than 1~nm and M=N=3 for diameters $D$ larger than 1~nm. In the case of bulk zinc-blende structures the minimization of the total energy leads to
theoretical cubic lattice constants $a_0=5.61$ {\AA} (GaAs), 5.83
{\AA} (InP), and 6.03 {\AA} (InAs) as well as to negative cohesive energies per cation-anion pair (without spin-polarisation corrections)
$\mu^{\rm bulk}_{III-V}=-9.68$ eV (GaAs), --9.49 eV (InP), and
--8.88 eV (InAs).

Free clean and passivated surfaces of $zb$ or $w$ crystals are
studied within the repeated-slab method using material slabs of about 14 to 17~{\AA} separated by a vacuum region of the same thickness~\cite{15}. 
The material slabs usually represent multiples of irreducible slabs. For the \{$1\bar 10$\}\emph{zb} surface we use four irreducible slabs each of them consisting of two atomic layers. The \{$11\bar 2$\}\emph{zb} surface is described using three irreducible slabs with six atomic layers. All considered surfaces are stoichiometric and electrostatic neutral. Nevertheless, in the case of \{$11\bar2$\} facets the different numbers of dangling bonds per anion or cation give rise to two unequal surfaces. In these cases we can only calculate an averaged value for the two surface energies. Their application to the nanorod energetics does, however, not influence the results since such inequivalent surfaces occur only pairwise in form of opposite facets on a nanorod. To model the wurtzite surfaces in the same manner we apply 11 atomic layers for the \{$1\bar 100$\}\emph{w} surfaces and seven atomic layers for the \{$11\bar 20$\}\emph{w} surfaces in the slab. All calculations are performed using a 1x1 translational symmetry. A possible surface reconstruction is not taken into account.

\subsection{Nanorod structure}

The one-dimensional translational symmetry in growth
direction [111]/[0001] is given by irreducible slabs of six atomic layers
(three cation-anion bilayers) of $zb$ in [111] direction or of
four atomic layers (two cation-anion bilayers) of $w$ in [0001]
direction~\cite{15}. 
Therefore, sometimes one speaks about the 3C(\emph{zb}) and 2H(\emph{w}) polytype of a compound. 
In our simulations we use slabs of the length  $L=2\sqrt{3}a_0$ of two (three) irreducible $zb$ ($w$) slabs. 
The diameter of the rod is characterized by the distance $D$ between opposite surface facets (see Fig.1).
Nanorods with hexagonal cross-sections and two different stackings ($zb$ or $w$)
of bilayers in growth direction are studied as observed in most of the experimental studies of III-V semiconductors~\cite{8,10,15b}. Therefore we restrict ourselves to those rods, where the point-group symmetry
$C_{3v}$ is conserved for arrangements of both types of stackings as in the $zb$- and
$w$-structures. We do not consider nanowires with high-index side facets, because of their roughness and possible higher dangling-bond densities. Surface normal orientations are studied only parallel to high symmetry directions. 
Hence, we end up with two different types I and II
of side facets. Type-I wires contain
$\{11\bar{2}\}$ and $\{1\bar{1}00\}$ side facets in the $zb$ and $w$
case, respectively (Fig.~2, left). GaAs rods with such surfaces can, e.g., be grown by MOVPE~\cite{3}. In
the type-II case, $\{1\bar{1}0\}$ facets for $zb$ and
$\{11\bar{2}0\}$ facets for $w$ occur as also observed experimentally (Fig.~2, right).~\cite{7,10,Mandl}   
 
\begin{figure}
\includegraphics[height=10cm]{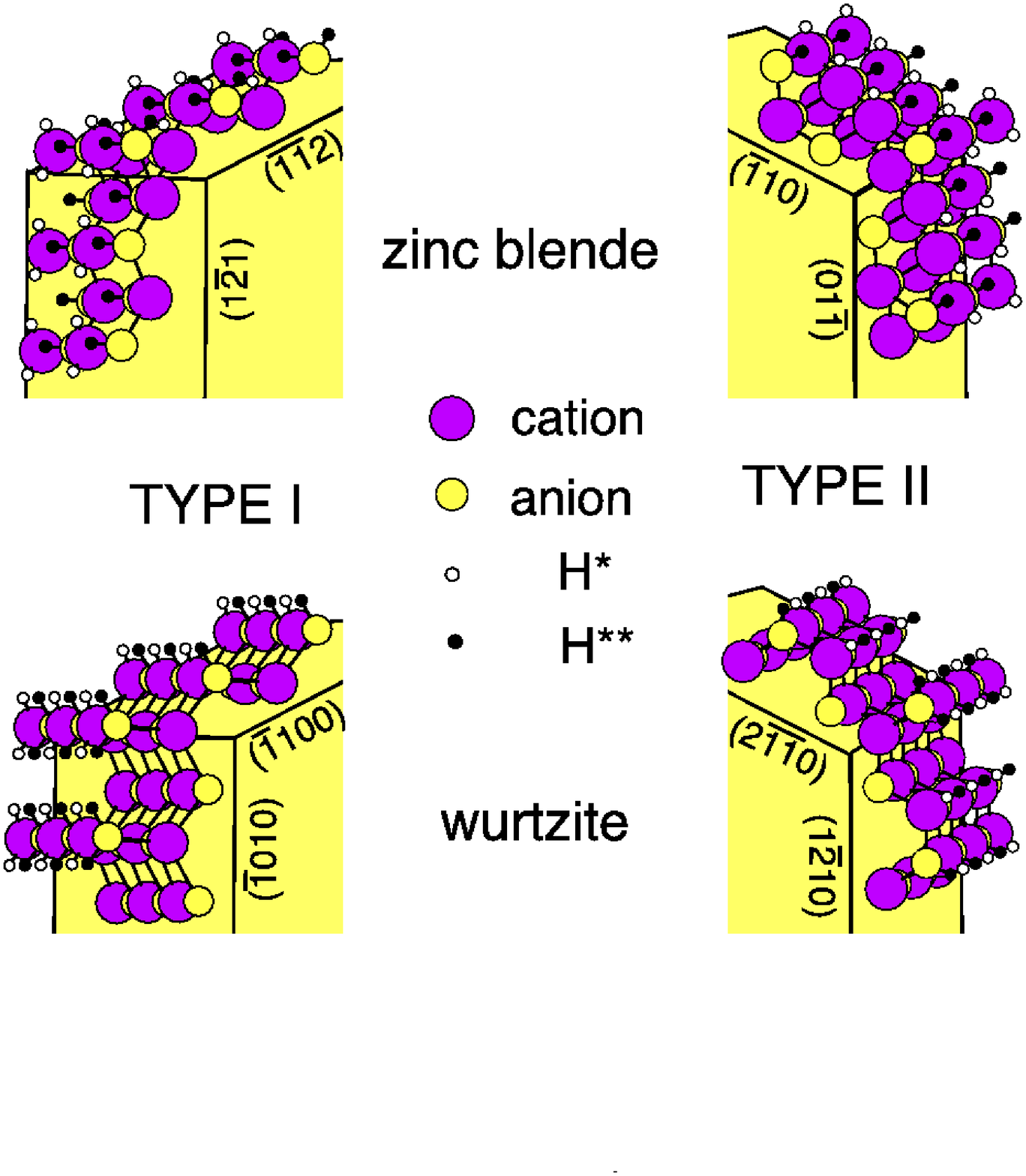}
\vspace{-2cm}
\caption{(Color online) Schematic stick and ball representation of adjacent facets on
type-I (left-hand side) and type-II (right-hand side) nanowires.}
\end{figure} 
 
To construct the nanowires on an atomic scale we start from a cation-anion bond
parallel to the growth direction in the center of the nanorod. The nearest-neighbor atoms occur at the corners of a tetrahedron along cubic body diagonals. In the wurtzite case each second tetrahedron in stacking direction is
twisted by 180$^\circ$ resulting in a shortening of the period of
translational symmetry in [0001] growth direction to two cation-anion bilayers compared to the three bilayers in the zinc-blende case.

\begin{figure*}
\includegraphics[height=8cm]{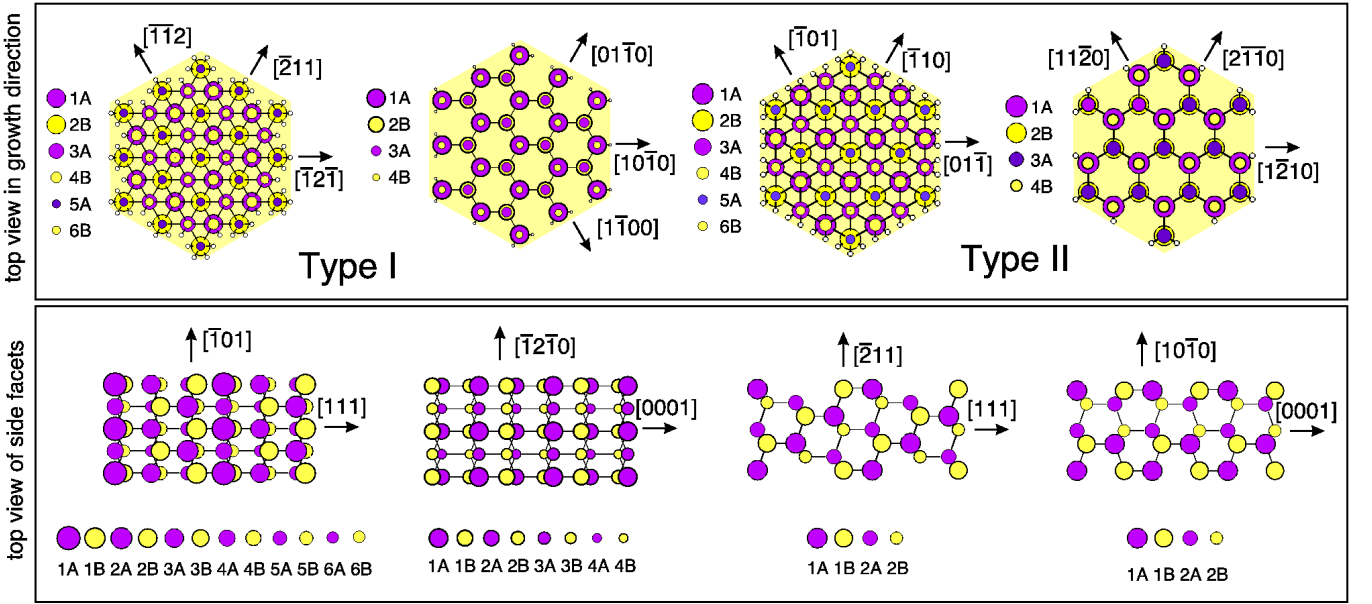}
\caption{(Color online) Schematic stick and ball representation of
type-I (left-hand side) and type-II (right-hand side) nanowire models.
In the case of the top views (upper panels) only the atoms in the
irreducible slab in different atomic layers are shown. The top views of the corresponding surface facets (lower panel) represent one possible arrangement versus the
length of two (three) irreducible $zb$ ($w$)
slabs. The circle size indicates atoms in different atomic layers 1-6, while the circle color represents cations~(A) or anions~(B). Pseudohydrogen atoms are represented by very small circles.}
\end{figure*}

The side facets (top view) of the two wire types considered are also shown in the lower panel of Fig.~3. They are stoichiometric and electrostatically neutral. Even the numbers of
cation and anion dangling bonds (DBs) are equal in each
irreducible slab. The absolute DB densities of the corresponding
infinite surface facets (i.e., neglecting the edge influence) vary
between $2\sqrt{2}/a_0^2$ for type-II $zb$- and $w$-wires, $4\sqrt{2}/[a_0^2\sqrt{3}]$
for type-I $zb$-wires, and $3\sqrt{6}/[2a_0^2]$ for type-I $w$-wires assuming the same bonding tetrahedra for \emph{w} stacking with ideal hexagonal lattice constants $a=a_0/\sqrt{2}$ and $c=2a_0/\sqrt{3}$. 
In the latter case the DB densities vary with the termination of the surface. Since the considered rods always exhibit both types of terminations at opposite facets we have just given the averaged value of both DB densities. To simulate a possible passivation of the DBs
at the wire surfaces due to adsorption of environmental atoms or
metal atom excess as schematically shown in Fig.~1, we use fractionally charged pseudohydrogen atoms with
a valence charge $Z=0.75$ $({\rm H}^*)$ or $1.25$ $({\rm H}^{**})$
\cite{16}. To obtain the equilibrium positions of the passivating
pseudoatoms we apply a conjugate gradient algorithm based on
\emph{first principles} forces. 

Figures 2 and 3 indicate remarkably different
geometries of the two types of side facets. In the type-II case the
side facets have only one DB per surface atom and exhibit rather
similar geometries. For
\{1$\bar{1}$0\}$zb$ facets the cation-anion bonds form a staircase-like
structure, while the twist of the bonding tetrahedra in the
\{2$\bar{1}\bar{1}$0\}$w$ case gives rise to zigzag chains in [0001]
direction. In contrast, type-I side facets differ remarkable in
their bonding structures. In the average, each surface atom possesses 4/3~($zb$) or 3/2~($w$)
DBs. However, already for $zb$ stacking their distribution depends
on the orientation of a $\{11\bar{2}\}$ facet. Neighboring facets,
for instance ($\bar{1}\bar{1}2$) and $(\bar{2}11)$, are different
with respect to their DB distribution as indicated in Fig.~2. On the surface only cation-anion bonds parallel to [0001]/[111] occur. The atoms in the uppermost
atomic layer possess an averaged value of 1.5 DBs for both crystal structures. However, the atoms in the second atomic layer of the (11$\bar{2}$)$zb$
surface exhibit one additional dangling bond. In
principle, the situation in nanorods is somewhat more complicated because of
the existence of edges which modify the picture of free surfaces
for finite wire sizes. However, for thick rods, i.e., in the limit
$ D \rightarrow\infty$, the facets represent free surfaces.

\subsection{Formation energy of a nanorod}

In order to study the stability of the different nanowires we investigate the formation energy of a nanorod $E^{\rm rod}_{\rm surf}$ with a certain surface area consisting of
symmetry-equivalent facets defined by diameter $D$ and length $L$ (defined in Fig.~1).~\cite{15} 
Taking into account that in a rod not all group-III and
group-V atoms are fourfold coordinated but the cation and anion
DBs occur pairwise the formation energy obeys the expression 
\begin{eqnarray}
E^{\rm rod}_{\rm form}&=&E^{\rm rod}_{\rm tot}(N_{III-V},N_{\rm
mol})- N_{III-V}\mu^{\rm bulk}_{III-V}\nonumber\\ &-&N_{\rm mol}\mu^{\rm
mol}_{H^{*}-H^{**}}-N_{\rm mol}\Delta\mu_{H^{*}-H^{**}},
\end{eqnarray}
where $E^{\rm rod}_{\rm tot}(N_{III-V},N_{\rm mol})$ is the total
energy of a rod containing $N_{III-V}$ cation-anion pairs with DBs 
or passivated by $N_{\rm mol}$ pairs of
pseudohydrogens H$^*$ and H$^{**}$. This energy is related to the corresponding bulk energy 
$N_{III-V}\mu^{\rm bulk}_{III-V}$ and to
the energy of the $N_{\rm mol}$ ${\rm H}^*-{\rm H}^{**}$ molecules
used to passivate the total number of $2N_{\rm mol}$ DBs. We assume that the rods are in thermodynamic equilibrium with a $zb$ III-V crystal that represents a reservoir for group-III and group-V atoms. The
passivation is governed by a reservoir of ${\rm H}^*-{\rm
H}^{**}$ molecules with the chemical potential
$\mu_{H^{*}-H^{**}}$. The variation of this chemical potential
allows to model the environmental influence during the growth
or annealing process as have been demonstrated for individual surfaces.~\cite{17} Here $\mu^{\rm mol}_{H^{*}-H^{**}}$ (calculated
as the negative molecule binding energy) corresponds to the situation where
the rod surface is exposed to molecular pseudohydrogen at
vanishing temperature. However, due to the need to overcome the
dissociation barrier, the stable passivated surface structures
will not form immediately. 
The surface coverage depends on the environmental conditions or the partial pressures and temperatures in the regions. This
dependence $\Delta\mu_{H^{*}-H^{**}}$ of $\mu_{H^{*}-H^{**}}$ can
be approximated similarly to that of a two-atomic ideal gas. A
slightly negative potential (e.g.
$\Delta\mu_{H^{*}-H^{**}}\approx-1$ eV in the case of true
hydrogen) may model typical MOVPE growth conditions while fully
passivated surfaces may form for $\Delta\mu_{H^{*}-H^{**}}>0$
where atomic pseudohydrogen is available.~\cite{17}
In a real preparation process, e.g. for oxidation of rod surfaces in air, the adsorption processes are more complex. In addition modifications of the surface stoichiometry may appear. Nevertheless, the study of the rod formation energy (1), which is based on Kramer's grand canonical potential, should give important information about the equilibrium state of rods in a certain gaseous environment. A generalization of (1) towards individual group-III or group-V atoms is possible by introducing the corresponding chemical potentials, in order to model different ratios of these species during the growth process.~\cite{15,F1,F2} 

The formation energy (1) depends on the geometry (diameter $D$, length
$L$) of the considered rod and the preparation conditions, i.e.,
$\Delta\mu_{H^{*}-H^{**}}$. We refer
this energy to the surface area $A_{\rm rod}=2\sqrt{3}\;DL$ of the considered rod and set
$\Delta\mu_{H^{*}-H^{**}}\equiv 0$. The resulting surface free
energy per unit area \cite{18,19}
\begin{eqnarray}
\tilde{\gamma}_{N_d}&=&\big[E^{\rm rod}_{\rm tot}(N_{III-V},N_{\rm mol})-
N_{III-V}\mu^{\rm bulk}_{III-V}\nonumber\\ & & \;\; - N_{\rm mol}\mu^{\rm
mol}_{H^{*}-H^{**}}\big]/ A_{\rm rod}
\end{eqnarray}
is influenced by the energetics of the inner rod atoms as will be discussed later, the energetics on a given surface facet, 
and the energetics of the edges between two facets. 
It still depends on the thickness of the rod, which determines the relative contribution of the three energies. Instead by $D$, in (2) we characterize the thickness by the dimensionless parameter $N_d$:
\begin{eqnarray}
N_d & = & \sqrt{{8 \over 3}}{D \over a_0 }\;\;\; \mbox{(type I rods),}\\
N_d & = & \sqrt{2}{D \over a_0 } \;\;\; \mbox{(type II rods),}
\end{eqnarray}
in terms of the ideal lateral lattice constant $a_0/\sqrt2$. The factor $2/\sqrt3$ is related to the 30$^{\text{o}}$ rotation between both types of rods.

\section{Results and Discussion}
\subsection{Energetics and phase diagram}

For $N_d\rightarrow\infty$ the
energy (2) converges against the surface energy $\tilde{\gamma}=\tilde{\gamma}_{\infty}$
of the clean or passivated surface with considered orientation as demonstrated in Table~I. 
Reconstruction or relaxation of the surfaces are not taken into account.~\footnote{Only the positions of the passivating pseudohydrogen atoms are relaxed until the Hellmann-Feynman forces are less than 20 meV/$\AA$.} 
The convergence is faster for passivated rods than for rods with clean facets. 
For the $w$-crystal structure $\tilde{\gamma}_{N_d}$~(2) is not exactly equal to the free surface energy. It contains contributions resulting from the change of the crystal structure. However, 
because of the small difference in the cohesive energies, e.g. $\mu_{III-V}^{\rm bulk} (w) - \mu_{III-V}^{\rm bulk} (zb) \approx 25$~meV for InAs, 
those contributions are very small and can therefore be neglected (at least for the considered numbers $ N_{III-V}$). Our surface results are in good agreement with other \emph{first principles} calculations. For instance the free surface energy of the relaxed cleavage (110) surface of InAs with
the value of 40.5~meV/\AA$^2$ agrees well with the result of Ref.~\onlinecite{19}.

Table I exhibits three clear trends. First,
the surface energies of the rods are only slightly larger than
the one of the corresponding free surface. With rising rod diameter,
$N_d\rightarrow\infty$, a rapid convergence to the free surface
values is observed. 
Already for InAs nanorods with diameters of about 1.7~nm ($N_d$=4, type II) the results are only slightly above those for very thick rods. 
Hence, the diameter of the rod is of little influence
on the stability of the rod, taking as measure the minimum surface
energy with respect to the surface area $A_{\rm rod}$ for given
volume \cite{20}. Second, for the assumed preparation conditions,
$\Delta\mu_{H^{*}-H^{**}}=0$, the energies of the passivated
surfaces are considerably smaller, indicating the stability of the
passivated rods versus such with clean surfaces. Third, the type-I
rods with a higher dangling-bond density of the surface possess a
larger surface energy independent of the rod diameter and the passivation. All values in Table~I follow the sequence of DB densities: type~II(\emph{zb})$\approx$type~II(\emph{w}) $<$ type~I(\emph{zb}) $<$ type~I(\emph{w}).
A clear favorization of a certain stacking with a certain diameter, as recently reported by means of total energy calculations using empirical pseudopotentials,~\cite{I} cannot be concluded.  
Which stacking order, $zb$ or $w$, in the rod is
more stable depends on the passivation and the wire type.

\begin{table}[!htb]
\caption{Surface energies $\tilde{\gamma}_{N_d}$ (in meV/\AA$^2$) for InAs
rods of type I or II for varying diameter $\sim N_d$. The values
for $N_d=\infty$ are calculated for free surfaces using the
repeated-slab method. The first (second) value has been obtained
for clean (passivated) nanorods.}
\begin{ruledtabular}
\begin{tabular}{|c|c|c|c|c|}
$N_d$ & \multicolumn{2}{c|}{Type I} & \multicolumn{2}{c|}{Type II}
\\ \cline{2-5}
 & $zb$ & $w$ & $zb$ & $w$ \\ \hline
 2 & 95 & 110 & 85 & 73 \\
 & 31  & 37 & 28 & 27 \\
 3 & -- & -- & 79 & 81 \\
   & -- & -- & 26   & 28 \\
 4 & 87 & 97 & -- & -- \\
   & 29    & 33    & -- & -- \\
$\infty$ & 82 & 84 & 65 & 63 \\
   & 27 & 33 & 23 & 25 \\ \hline
\end{tabular}
\end{ruledtabular}
\end{table}

The stability of InAs rods with different stacking, surface
orientation, and surface passivation as a function of the
preparation conditions is demonstrated by the phase diagram in
Fig.~4. According to a Wulff-like approach~\cite{20,G} the most stable rod should have the lowest surface energy $\gamma_{N_d}=E^{\rm rod}_{\rm surf}/A_{\rm rod}$. This allows to identify the most favorable facet orientation, surface coverage, and bilayer stacking for given preparation conditions and fixed diameter. 
We have plotted the surface energy $\gamma_{N_d}$ versus the variation of the chemical potential of the passivating species. 
Negative values of $\Delta\mu_{H^{*}-H^{**}}$ stand for passivating species-poor conditions, while positive values indicate the availability of free passivating atoms. 
Because of the weak diameter dependence of the energies $\gamma_{N_d}$ we use the free surface values $\gamma_{\infty}$ scaled with the corresponding DB densities, which are diameter dependent. In this way we are able to extrapolated our results to rods with a realistic diameter of about 5~nm.   
One observes a strong
influence of the preparation conditions on the rod stability. If mainly
passivating molecules of a not too high density are present, the
rods prefer clean surfaces of type II with low dangling-bond
density, i.e., $\{1\bar{1}0\}/\{2\bar{1}\bar{1}0\}$ facets. In the
presence of dissociated molecules (free radicals H$^*$ and
H$^{**}$), i.e., $\Delta\mu_{H^{*}-H^{**}} > 1$~eV , the rod should be passivated and possesses type-I facets,
i.e., those with originally higher dangling-bond densities. In this case, a clear favorization of the wurtzite stacking
versus the zinc-blende stacking can be stated. However, in all other cases the
energies of the two different bilayer stackings are very close, so that probably other (e.g. kinetic) aspects of the rod
growth are more important for a given stacking than energies derived
for pure equilibrium situations.
Under intermediate conditions, i.e., $\Delta\mu_{H^{*}-H^{**}}$ between -1~eV and zero, passivated type-II facets are preferred. Although type-II facets have the same dangling-bond density for $zb$ and $w$ stacking, we find a slight preference of the $zb$ structure. This is a consequence of the higher negative bulk cohesive energy $\mu^{\rm bulk}_{III-V}(w)$ of the $w$ structure and the influence of edges on the DBs.
The experimental results
concerning InAs are usually not unique.~\cite{6,4} 
However, in Refs.~\onlinecite{4} and~\onlinecite{Mandl} both types of stacking, \emph{zb} and \emph{w}, have been observed under MOVPE conditions, which is in agreement with our predictions.

\begin{table}[!htb]
\caption{Surface energies $\tilde{\gamma}_{N_d}$ (in meV/\AA$^2$) for passivated GaAs, InP, and InAs rods of type I/II for varying diameter $\sim N_d$. The values
for $N_d=\infty$ are calculated for free surfaces using the
repeated-slab method.}
\begin{ruledtabular}
\begin{tabular}{|c|c|c|c|c|c|c|}
$N_d$ & \multicolumn{2}{c|}{GaAs} & \multicolumn{2}{c|}{InP} &
\multicolumn{2}{c|}{InAs} \\
 & $zb$ & $w$ & $zb$ & $w$ & $zb$ & $w$ \\ \hline
 2 & 25/21 & 30/22 & 26/22 & 30/21 & 31/28 & 37/27 \\
 4/3 & 23/19 & 28/22 & 24/20 & 28/22 & 29/26 & 33/28 \\
$\infty$ & 21/16 & 29/19 & 24/17 & 30/19 & 27/23 & 33/25 \\ \hline
\end{tabular}
\end{ruledtabular}
\end{table}

In order to study chemical
trends within the III-V semiconductors GaAs, InP, and InAs with $zb$ bulk ground states, we
restrict the computations to passivated rods according to the wide stability range in the InAs phase diagram in Fig.~4. 
The energy variation with the compound may be larger for clean surfaces. Nevertheless, they should only occur under ultra-high vacuum conditions. Results for $\tilde{\gamma}_{N_d}$ are listed in Table II for different
wire thicknesses. 
The least(most) stable passivated nanorods are obtained for InAs(GaAs). 
However, the most important consequence of Table~II is, that the surface energies of all considered semiconductors show the same behavior and only a small reduction from InAs via InP to GaAs. Therfore the phase diagrams of the considered III-V semiconductor nanorods will have the same qualitative shape as that of InAs in Fig.~4 Hence, the occurrence of both stacking types, wurtzite and zinc blende, in III-V semiconductor nanorods may be explained by the weak dependence of the energy sequence on
the bilayer stacking and the rod diameter (as discussed in detail for InAs).

\begin{figure}[!htb]
\vspace{0.5cm}
\includegraphics[height=6.2cm]{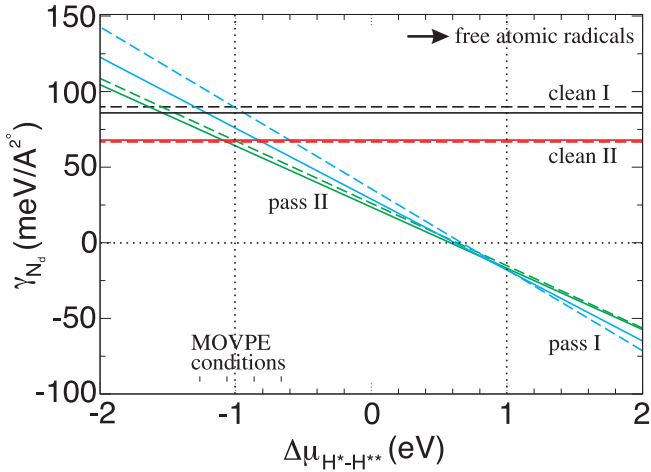}
\caption{(Color online) Surface energy per surface area $\gamma_{N_d}$
of a InAs rod with a thickness of about 5 nm versus the variation of the chemical potential
$\Delta\mu_{H^{*}-H^{**}}$ of the passivating molecules. 
Solid lines: $zb$ stacking, dashed lines: $w$ stacking. Rod type and surface coverage are indicated.
The central part surrounded by thin dotted lines describes the region of the most reliable chemical potentials.}
\end{figure}

\subsection{Geometry and bond lengths}

All results presented until now have been computed without lattice relaxation. The atomic positions were obtained from
the tetrahedral structure and the (theoretical) bulk cubic lattice
constant $a_0$. Only the stacking has been changed. 
In order to study the influence of atomic rearrangements due to the surface presence, we allow atomic relaxation in the nanorods under the constraint of conservation of the tetrahedral coordination. Since both stackings,
$zb$ (3C) and $w$ (2H), can be described in hexagonal unit cells
with lattice constants $a$ and $c$, we allow for changes of these
lattice constants with respect to the corresponding ideal bulk values
$a=a_0/\sqrt{2}$ ($zb$, $w$), $c=\sqrt{3}a_0$ $(zb)$ or
$c=\frac{2}{\sqrt{3}}a_0$ ($w$). 
For thin rods ($N_d$=2) with clean side facets  
we find a nearly vanishing increase of the $a$ lattice constant by 0.06~\% for both stackings. 
Contrariwise the $c$ lattice constant is reduced by 1.7~\% ($zb$) or 2.8~\% ($w$). As thermodynamically expected this leads to a small shrinkage of the III-V bond lengths in the rods. 
Surface tension is accompanied by a tendency for volume reduction. A similar tendency has been observed for semiconductor nanocrystals.~\cite{H} The absolute surface energies $\gamma_{N_d}$ are somewhat changed with respect to the unrelaxed rod, but their relative differences remain almost the same. 
With the $c$ lattice constant the cation-anion bilayer thickness parallel to the growth direction is decreased. 

For passivated rods the already small changes nearly vanish, i.e., the $a$ lattice constant remains constant, while the $c$ lattice constant is reduced by less than 0.8~\% for both stacking types.  
This observation is in good agreement with the almost bulk-like bilayer thicknesses measured by high-resolution TEM~\cite{5,6,10}. 
For $zb$ GaAs nanorods only a tiny reduction (less than the error bar) of the lattice plane spacing in growth direction was found in Refs.~\onlinecite{5,6}, while for $w$-InP nanorods a reduction of the $c$ lattice constant by 2.8\% was observed. However, we cannot confirm the increase of the $c/a$ ratio with respect to the ideal value, which was found in Refs.~\onlinecite{Mandl} and~\onlinecite{Mandl19}.  

\section{Summary and Conclusions}

Using \emph{ab initio} total-energy calculations for thin nanorods and individual surfaces of the same materials we have studied the energetics of III-V semiconductor nanorods in thermodynamic equilibrium. For rod diameters smaller than two nm formation energies are directly calculated for a given segment length. For thick rods the rod surface energy is obtained from values of the individual facets using a Wulff-like construction. The preparation conditions, especially the presence of passivating species, are simulated by corresponding chemical potentials (changing from the free energy to the grand canonical thermodynamic potential). We studied rods grown in [111]/[0001] direction with varying diameter, bilayer stacking, surface coverage, and facet orientation. For a given chemical potential of the passivating species the preference of a certain rod type roughly follows the trend that the surface energy of the wire crystal is lowered by decreasing the density of surface dangling bonds. However, for extreme preparation conditions, e.g. in the presence of reactive atoms this route can be violated and passivated structures with complete adsorbate overlayers are preferred.
    
In conclusion, we have shown that the rod energetics near the thermodynamic equilibrium is
ruled by their surfaces. The formation energies of those rods
approach the energies of the corresponding free
surfaces for large diameters.
The orientation of the surface facets and the
passivation of their dangling bonds play an important role, in
contrast to the bilayer stacking in growth direction, which only leads to minor energy variations. Our studies
are able to explain qualitatively a wide variety of results concerning bilayer
stacking and rod facets found for III-V semiconductors. At least clear trends can be derived for thermodynamic equilibrium conditions. In addition, the growth kinetics and the catalyst, the growth conditions and after-growth treatment, e.g. annealing and the corresponding atmosphere, may have an influence on the resulting nanorod.

\begin{acknowledgments}
We gratefully acknowledge valuable discussions with B.
Mandl and G. Bauer from the University of Linz. Financial
support was provided by the Fonds zur F\"orderung der
Wissenschaftlichen Forschung (Austria) via SFB25 IR-ON and the EU
NoE NANOQUANTA (Project No. NMP4-CT-2004-500198). We thank the
H\"ochstleistungsrechenzentrum Stuttgart for granted computer
time.
\end{acknowledgments}

\end{document}